\documentclass[11pt,a4paper]{article}
\usepackage{jcappub}
\usepackage[utf8]{inputenc}

\title{A local self-tuning mechanism for the cosmological constant}

\author[a]{Daniel Sobral Blanco}
\author[a]{and Lucas Lombriser}
\date{January 2020}

\affiliation[a]{D\'epartement de Physique Th\'eorique, Universit\'e de Gen\`eve,\\ 24 quai Ernest Ansermet, 1211 Gen\`eve 4, Switzerland}

\emailAdd{Daniel.Sobral@etu.unige.ch}
\emailAdd{Lucas.Lombriser@unige.ch}

\abstract{

Recently the global variation of the Planck mass in the General Relativistic Einstein-Hilbert action was proposed as a self-tuning mechanism of the cosmological constant preventing vacuum energy from freely gravitating.
We show that this global mechanism emerges for generic local scalar-tensor theories with additional coupling of the scalar field to the field strength of a three-form gauge field that turns the scalar field constant on the domain of the action.
Evaluation of the resulting integral constraint equation over the observable Universe yields a self-consistent framework with General Relativistic field equations and arbitrary radiatively stable residual cosmological constant.
We argue that the expectation value for this residual is in good agreement with the magnitude of the observed cosmic acceleration.
}

\newcommand{\integral}{\int_{\mathcal{M}}\,d^4x\sqrt{-g}\,}
\newcommand{\average}[1]{\left\langle #1 \right\rangle}
\newcommand{\vac}{\mathrm{vac}}
\newcommand{\planck}[1]{M_P^{\,#1}}
\newcommand{\g}{g_{\mu\nu}}
\newcommand{\residual}{\Lambda_\mathrm{res}}

\begin{document}

\maketitle

\section{Introduction}

Unravelling the underlying nature of the cosmological constant in Einstein’s Theory of General Relativity (GR) remains a persistent enigma to modern physics.
It is generally anticipated to represent the vacuum energy contribution to gravitational dynamics, which should be of adequate magnitude to account for the observed late-time accelerated expansion of our Universe~\cite{Riess:1998cb,Perlmutter:1998np}.
Quantum theoretical expectations for this contribution, however, exceed the measurement by $\gtrsim50$ orders of magnitude~\cite{Weinberg:1988cp,Martin:2012bt}.
While this may imply a missing prescription for the correct computation of standard vacuum energy contributions, it has also motivated the conjecture that vacuum energy may be prevented from gravitating to full extent by an undetermined mechanism~\cite{unruh,henneaux2,henneaux,barrow1,barrow2,kalpa4,kalpa1,lomb2,lomb1} and that cosmic acceleration could instead be due to a dark energy field permeating the Cosmos or a breakdown of GR at large scales~\cite{Koyama:2015vza,Joyce:2016vqv,Ishak:2018his}.
The dynamics of dark energy however must be fine-tuned to closely mimic a cosmological constant~\cite{Aghanim:2018eyx}, and the confirmed equality between the speeds of gravity and light~\cite{Monitor:2017mdv} combined with observations of the large-scale structure poses hard challenges to the concept of cosmic self-acceleration from a genuine modification of gravity~\cite{Lombriser:2015sxa,Lombriser:2016yzn}.

Recently, in Ref.~\cite{lomb1} the cosmological constant problem was re-examined under the aspect of an additional variation of the Einstein-Hilbert action of GR with respect to the Planck mass, performed along with the metric variation.
An interpretation of this approach is offered by the treatment of the Planck mass in the action as a global Lagrange multiplier that imposes GR dynamics on the metric prescribing the spacetime for the matter fields.
The resulting additional constraint equation prevents vacuum energy from fully gravitating.
Moreover, the evaluation of this constraint under consideration of the evolution of the inhomogeneous cosmic matter distribution was shown to self-consistently reproduce the observed cosmological constant with an expected value for its current energy density parameter of $\Omega_{\Lambda}=0.704$~\cite{lomb1}, in good agreement with current measurement~\cite{Aghanim:2018eyx}. Besides the non-gravitating vacuum, the additional Planck mass variation therefore also explains the rise of the late-time accelerated cosmic expansion and the coincidence of $\Omega_{\Lambda}$ with the current energy density of matter $\Omega_m$, also known as the \emph{Why Now?}~problem.

In this paper, we develop a local theory from which the global self-tuning mechanism obtained from the Planck mass variation naturally emerges.
To achieve this, we consider general scalar-tensor theories arising as the effective limit of a more fundamental theory with the addition of a topological sector to the action, in which the scalar field couples to the field strength of a three-form gauge field.
The resulting additional field equations enforce constancy of the scalar field on the domain of the action, and we discuss how this reproduces the global mechanism.

The paper is organised as follows.
Sec.~\ref{sec2} briefly reviews the global self-tuning mechanism from the global Planck mass variation of Ref.~\cite{lomb1}.
In Sec.~\ref{sec3} a local theory is developed based on generic scalar-tensor theories with additional coupling of the scalar field to the field strength of a three-form gauge field, from which the global mechanism emerges.
Sec.~\ref{sec4} discusses likelihood considerations for the value of the residual cosmological constant produced by the self-tuning mechanism.
Finally, we conclude in Sec.~\ref{sec5} and discuss some general aspects of the mechanism in Appendix~\ref{appendix1} and graviton loops in Appendix~\ref{appendix2}.

\section{The self-tuning of $\Lambda$ from a global Planck mass variation}\label{sec2}

Recently, the variation of the General Relativistic (GR) Einstein-Hilbert action with respect to the quadratic Planck mass in addition to the usual metric variation has been proposed as a self-tuning mechanism for the cosmological constant~\cite{lomb1}.
The approach allows for an interpretation of the Planck mass as a global Lagrange multiplier that imposes GR dynamics for the metric describing the geodesics of the matter fields of a given matter Lagrangian.
The two variations result in the usual Einstein field equations and an additional integral constraint equation that acts to self-tune the cosmological constant and prevents vacuum energy from freely gravitating.
We shall briefly review this global mechanism before discussing in Sec.~\ref{sec3} how it can emerge in generic local scalar-tensor theories with additional coupling of the scalar field to the field strength of a three-form gauge field.

Consider the Einstein-Hilbert action of GR,
\begin{equation}
\label{eq:21}
    S = \frac{\planck{2}}{2}\integral(R-2\Lambda) + \integral\mathcal{L}(g^{\mu\nu},\Psi_m) + \mathrm{b.t.} \,,
\end{equation}
where $\mathcal{M}$ denotes the cosmic manifold, $\Lambda$ is a free classical cosmological constant and $\mathrm{b.t.}$ refers to the Gibbons-Hawking-York boundary term.
Variation of the action~\eqref{eq:21} with respect to the metric $\g$ yields the Einstein field equations
\begin{equation}
\label{eq:22}
    G_{\mu\nu} + \Lambda\g = \planck{-2}\,T_{\mu\nu}
\end{equation}
where $T_{\mu\nu} \equiv -2\left[\delta\left(\sqrt{-g}\mathcal{L}_m\right)/\delta\g\right]/\sqrt{-g}$ denotes the energy-momentum tensor.
Following Ref.~\cite{lomb1}, in addition to the metric variation, we shall now perform a variation of the action~\eqref{eq:21} with respect to the quadratic Planck mass $\planck{2}$, where boundary conditions may be adapted as in Ref.~\cite{kalpa2} (also see Ref.~\cite{beltran}) and we will henceforth neglect the boundary term. To illustrate the cancellation of the vacuum and bare cosmological constants, $\Lambda_\vac$ and $\Lambda_B$, or rather their absorption in the self-tuning of the classical $\Lambda$, we shall first assume that they are independent of $\planck{2}$.
Hence, we assume the simple scaling of the vacuum contribution as $\planck{2}\Lambda_\vac \propto \planck{2}M^2$ for some renormalization mass $M$ (e.g., the leading-order behaviour found in Ref.~\cite{wang}). We now separate out the vacuum and bare components from the matter Lagrangian density, $\mathcal{L}_m = \Bar{\mathcal{L}}_m - \planck{2}(\Lambda_\vac + \Lambda_B)$, and vary the action~\eqref{eq:21} with respect to $\g$ and $\planck{2}$.
This gives the Einstein field equations
\begin{equation}
    \label{eq:24}
    G_{\mu\nu} + (\Lambda+\Lambda_\vac+\Lambda_B)\g = \planck{-2}\tau_{\mu\nu} \,,
\end{equation}
where $\tau_{\mu\nu}$ is specified by $\Bar{\mathcal{L}}_m$ and represents the stress-energy tensor of the usual matter components only.
The variation with respect to $\planck{2}$ yields the constraint
\begin{equation}
\label{eq:25}
    \integral\left[\frac{R}{2}-(\Lambda+\Lambda_\vac+\Lambda_B)\right] = 0 \,.
\end{equation}
Using the trace of Eq.~\eqref{eq:24} this implies that $\Lambda+\Lambda_\vac+\Lambda_B = \frac{\planck{2}}{2}\,\average{\tau}$, where $\average{\tau} \equiv \integral\,\tau/\integral$.
The constraint only needs to apply for a given choice, or measurement, of the Planck mass, hence, $\Lambda$ remains not explicitly dependent on $\planck{2}$ (see Sec.~\ref{sec3}).
The Einstein equations may therefore be written as
\begin{equation}
\label{eq:26}
   G_{\mu\nu} + \frac{\planck{-2}}{2}\average{\tau}\g = \planck{-2}\tau_{\mu\nu}
\end{equation}
and, hence, the vacuum and bare contributions to the cosmological constant do not freely gravitate.
For simplicity, in the following, we will restrict our discussion to the vacuum term only, but we will address details on the cancelling of $\Lambda_B$ in Appendix~\ref{appendix1}.

The result in Eq.~\eqref{eq:26} is reminiscent of vacuum energy sequestering~\cite{kalpa4}, but the cancellation of the problematic contributions occurs here in a different fashion. Rather than a cancellation between the left- and right-hand sides of the Einstein equations as in the sequestering framework, the value of $\Lambda$ is set here by the constraint equation~\eqref{eq:25} such that the sum of the cosmological constants must match the quantity $\planck{-2}\average{\tau}/2$, the residual, or effective, cosmological constant.
Interestingly, the same fraction was found to fix the cosmological constant in Ref.~\cite{gaztanaga} from the consideration of a boundary condition on the causal region around an observer.

So far, we have only considered the simple scaling $\planck{2}\Lambda_\vac \propto \planck{2}M^2$.
However, it is not granted that the vacuum contribution should scale as such. More generally, we may assume a power-law relation $\planck{2}\Lambda_\vac = \planck{2\alpha} \bar{\Lambda}_{\mathrm{vac}}$, where the bar denotes the Planck mass independent part.
To cancel this term, we also need a classical counter-term $\planck{2\alpha} \bar{\Lambda}_{\alpha}$.
With the same procedure as for Eq.~\eqref{eq:26} this yields the field equations~\cite{lomb1}
\begin{equation}
\label{eq:27}
    G_{\mu\nu}+\frac{1}{2-\alpha}\left[(1-\alpha)\Lambda+\frac{\planck{-2}}{2}\average{\tau}\right]\,\g = \planck{-2}\tau_{\mu\nu} \,,
\end{equation}
where $\Lambda$ remains a free classical cosmological constant that is radiatively stable and determined by measurement.
For $\alpha = 1$, Eq.~\eqref{eq:27} reduces to Eq.~\eqref{eq:26}.
For $\alpha = 0$, one recovers the dynamical equations of the local sequestering mechanism~\cite{kalpa1} with $\Lambda_{\mathrm{tot}} = \tfrac{1}{4}\planck{-2}\average{\tau} + \Delta\Lambda$, where $\Delta\Lambda = \Lambda/2$.

We can further relax the power-law assumption and consider a series expansion of $\Lambda_\vac$ in $\planck{2}$, for instance introduced by graviton loops~\cite{kalpa5}. We discuss this scenario in Appendix~\ref{appendix2}.
Similarly, if independent of Planck mass, quantum corrections with higher-derivative terms in Eq.~\eqref{eq:21} do not contribute to Eq.~\eqref{eq:25} or the field equations (also see Ref.~\cite{kalpa5}).
If dependent on $\planck{2}$, they are cancelled by the classical counter-term.
In the scalar-tensor representation discussed in Sec.~\ref{sec3}, a coupled Gauss-Bonnet invariant could also be recast as a Horndeski theory, for which the self-tuning is shown to work in Sec.~\ref{sec32}.
Importantly, we can even allow for arbitrary functions of the quadratic Planck mass for both $\Lambda_\vac(\planck{2})$ and $\Lambda_B(\planck{2})$. What is needed for the cancellation is the addition of a classical counter-term which is taken to be a free function of $\planck{2}$.
This recovers Eq.~\eqref{eq:27} with $\alpha=\partial\ln\Lambda_\vac/\partial\ln\planck{2}$ (see Appendix~\ref{appendix1}).

One may wonder about the fundamental nature giving rise to a global Planck mass variation of the Einstein-Hilbert action.
It is worth noting the similarity of this variation to a scalar-tensor theory in Jordan-Brans-Dicke representation with constant scalar field across the observable universe, and we shall explore this connection in more detail in Sec.~\ref{sec3}.
A transformation into Einstein frame then changes the variation from one in $\planck{2}$ to one with respect to an effective $\Lambda$ and a coupling in the matter sector.
The approach therefore shares similarities with the proposals of Refs.~\cite{unruh,henneaux2,henneaux,barrow1,barrow2} but it is also different as, for instance, it does not impose the constant four-volume of unimodular gravity.
We can exploit the similarities between these frameworks to address the question of how a the global Planck mass variation may arise from a local theory of gravity.
For example, the scalar field can become a spacetime constant when a $\delta$-function is generated through appropriate boundary conditions on an additional vector field~\cite{henneaux2,barrow1,barrow2}.
Alternatively, it can be turned constant through coupling it to an additional squared four-form field strength as can arise in supergravity~\cite{henneaux,barrow1,barrow2,aurilia,hawking,bousso}.
This approach has been adopted as well in the local sequestering framework~\cite{kalpa1}. One may also envisage a type II multiverse scenario, where different observable patches may be equipped with different Planck masses, which could be accompanied with a variational principle and formulated in terms of a partition function (cf.~\cite{barrow1,barrow2}).

In the following we will focus on the emergence of the global mechanism from generic scalar-tensor theories endowed with an additional coupling of the scalar field to the field strength of a three-form gauge field.

\section{A local theory} \label{sec3}

Having reviewed the self-tuning mechanism of the cosmological constant from the global Planck mass variation of the GR Einstein-Hilbert action in Sec.~\ref{sec2}, we shall now explore one of the candidates for a local theory that gives rise to this global mechanism.
We will put our focus on scalar-tensor theories.
In Sec.~\ref{sec31} we show how a simple scalar-tensor model with additional coupling of the scalar field to the field strength of a three-form gauge field will enforce constancy of the scalar field over the domain of the Einstein-Hilbert action and reproduce the global self-tuning mechanism.
We then show in Sec.~\ref{sec32} how this approach applies to the most general classes of scalar-tensor theories.
Finally, in Sec.~\ref{sec33} we will discuss the correspondence between the local and global mechanisms in more detail.

\subsection{Self-tuning mechanism for a simple scalar-tensor theory}\label{sec31}

Let us first consider the simple scalar-tensor action
\begin{equation}
\label{eq:31}
    S = \integral \left[\frac{1}{2}\varphi R - V(\varphi) + \mathcal{L}_m(\g,\Psi_m)\right] \,,
\end{equation}
which shall represent the effective limit of a more fundamental theory, for instance, obtained from the compactification of a higher-dimensional theory of gravity.
Note that for now we do not include a kinetic term.
Hence, Eq.~\eqref{eq:31} corresponds to a Jordan-Brans-Dicke action with Brans-Dicke parameter $\omega=0$, as is the case in $f(R)$ gravity.
GR is recovered in the limit of $\varphi\rightarrow\planck{2}$.
We again perform the separation $\mathcal{L}_m=\bar{\mathcal{L}}_m - \planck{2}\Lambda_{\mathrm{vac}}(\varphi)$, where $\Lambda_{\mathrm{vac}}(\varphi)$ shall be an arbitrary function of $\varphi$.
Recall that the bare contribution $\Lambda_B$ is discussed in Appendix~\ref{appendix1}.

In addition to the scalar-tensor action~\eqref{eq:31}, we shall introduce the topological contribution
 \begin{equation}
 \label{eq:32}
     S_{\mathrm{A}} = \frac{1}{4!}\int_{\mathcal{M}}d^4x\, \epsilon^{\mu\nu\rho\sigma}\,\sigma(\varphi)\,F_{\mu\nu\rho\sigma} \,,
 \end{equation}
where $F_{\mu\nu\rho\sigma}= \partial_{[\mu}A_{\nu\rho\sigma]}$ is the field strength of a three-form gauge field $A_{\nu\rho\sigma}$ coupled to the scalar field $\varphi$ through a function $\sigma(\varphi)$.
Note the similarity to the local sequestering framework~\cite{kalpa1} (also see Refs.~\cite{henneaux,barrow1,barrow2,aurilia,hawking,bousso}).
In contrast to the sequestering mechanism, however, we only have one scalar field and one topological sector in Eqs.~\eqref{eq:31} and \eqref{eq:32} since the potential $V(\varphi)$ is a function of the gravitational coupling $\varphi$. It is worth noting that $S_A$ is a term that is common to supergravity models~\cite{henneaux,barrow1,barrow2,aurilia,hawking,bousso}.
The role it plays here is to fix the dynamics of the scalar field $\varphi$ to take the constant value $\planck{2}$ across the spacetime $\mathcal{M}$.

We will now see that the local theory described by the total action $S+S_A$ reproduces the results of the global self-tuning mechanism discussed in Sec.~\ref{sec2}. Variation of the total action with respect to the metric yields the modified Einstein equations
\begin{equation}
 \label{eom7}
    \varphi G_{\mu\nu} + \left[V(\varphi)+\planck{2}\Lambda_{\mathrm{vac}}(\varphi)\right] g_{\mu\nu} = (\nabla_\mu\nabla_\nu-g_{\mu\nu}\Box) \varphi + \tau_{\mu\nu} \,,
\end{equation}
where $\tau_{\mu\nu}$ is again the stress-energy tensor specified by $\bar{\mathcal{L}}_m$.
Variation of the action with respect to $A_{\mu\nu\rho}$ gives the crucial condition
\begin{equation}
 \label{eq:constantphi}
 \partial_\mu \varphi = 0 \,.
\end{equation}
Thus, the dynamics of the scalar field is fixed in the sense that it does not have any propagating degrees of freedom or local fluctuating modes~\cite{henneaux2}.
Finally, varying the total action with respect to $\varphi$, one obtains the constraint equation
\begin{equation}
 \label{eq:constraint}
    \integral\left[\frac{1}{2} R - V'(\varphi) - \planck{2} \Lambda_{\mathrm{vac}}'(\varphi) + \frac{\sigma'(\varphi)}{4!}\frac{\varepsilon^{\mu\nu\rho\gamma}}{\sqrt{-g}} F_{\mu\nu\rho\gamma}\right] = 0 \,,
\end{equation}
where primes denote derivatives with respect to $\varphi$.

Taking the trace of Eq.~\eqref{eom7} and using Eq.~\eqref{eq:constantphi}, one finds $\varphi R = 4\left[V(\varphi) + \planck{2} \Lambda_{\mathrm{vac}}(\varphi)\right] - \tau$ such that Eq.~\eqref{eq:constraint} can be recast as
\begin{equation}
 \label{eq:constraint2}
    \integral\left[ \varphi^{-1} \left(2 - \partial_{\ln\varphi}\right)\left( V + \planck{2} \Lambda_{\mathrm{vac}}\right) - \frac{\tau}{2\varphi} + \frac{\sigma'}{4!}\frac{\varepsilon^{\mu\nu\rho\gamma}}{\sqrt{-g}} F_{\mu\nu\rho\gamma}\right] = 0 \,,
\end{equation}
For convenience, we define $\alpha\equiv\partial\ln\Lambda_{\mathrm{vac}}/\partial\ln\varphi$ and $\beta\equiv\partial\ln \Delta V/\partial\ln\varphi$, where $\Delta V=V-V_c$ and $V_c$ shall play the role of the classical counter-term to $\Lambda_{\mathrm{vac}}$ as in Sec.~\ref{sec2} with $\partial\ln V_c/\partial\ln\varphi = \alpha$.
Note that $\alpha$ and $\beta$ do not need to be constants.
Eq.~\eqref{eq:constraint2} becomes
\begin{equation}
    \integral \varphi^{-1} \left[ (2 - \beta) \Delta V + (2 - \alpha) \left( \planck{2} \Lambda_{\mathrm{vac}}+V_c \right) - \frac{\tau}{2} + \frac{\sigma'\varphi}{4!}\frac{\varepsilon^{\mu\nu\rho\gamma}}{\sqrt{-g}} F_{\mu\nu\rho\gamma}\right] = 0 \,.
\end{equation}
This implies the constraint
\begin{equation}
\label{constraint4}
    (2-\beta) \planck{-2} \Delta V + (2 -\alpha) \left(\Lambda_{\mathrm{vac}} + \planck{-2} V_c \right) = \frac{\planck{-2}}{2}\langle\tau\rangle + \Delta\Lambda \,,
\end{equation}
where we have defined
\begin{equation}
 \label{eq:DLambda}
    \frac{\planck{2}}{\varphi} \Delta\Lambda \equiv -\frac{\sigma'}{4!}\left\langle\frac{\varepsilon^{\mu\nu\rho\gamma}}{\sqrt{-g}} F_{\mu\nu\rho\gamma}\right\rangle = - \frac{\sigma'}{4!}\frac{\int d^4x \varepsilon^{\mu\nu\rho\gamma} F_{\mu\nu\rho\gamma}}{\int d^4x \sqrt{-g}} \,.
\end{equation}
Finally, with Eq.~\eqref{eom7}, utilising $\varphi=\planck{2}$, we obtain the Einstein field equations
\begin{equation}
 \label{eq:localeinstein}
 G_{\mu\nu} + \frac{1}{2-\alpha} \left[(\beta-\alpha)\planck{-2} \Delta V + \frac{\planck{-2}}{2}\langle\tau\rangle + \Delta\Lambda\right] g_{\mu\nu}  = \planck{-2} \tau_{\mu\nu} \,,
\end{equation}
where the vacuum term is prevented from freely gravitating and $\Lambda = \planck{-2}\Delta V$ is a free, radiatively stable classical cosmological constant to be determined by measurement.
Note that with $\beta=1$, we recover Eq.~\eqref{eq:27} of the global self-tuning mechanism.
In contrast to Eq.~\eqref{eq:27}, however, we also obtain the additional term $\Delta\Lambda$. Importantly, $\Delta\Lambda$ does not take the same form as in Refs.~\cite{kalpa1,kalpa5}, where the denominator in the expression equivalent to Eq.~\eqref{eq:DLambda}, similarly to the numerator, is given by the flux of a second three-form gauge field. In Eq.~\eqref{eq:DLambda} the denominator is instead the four-volume of the cosmic manifold. With the flux of the 3-form gauge field in the numerator being a finite, small, UV-stable quantity and assuming the Universe grows sufficiently old, it is natural to expect that $\Delta\Lambda \rightarrow 0$.
Recall, however, that a free classical cosmological constant is still present with $\Delta V$.

\subsection{Generalisation to Horndeski action} \label{sec32}

The discussion and results presented in Sec.~\ref{sec31} can easily be generalised to broader classes of scalar-tensor theories such as Horndeski gravity, which describes the most general local scalar-tensor theory in four dimensions that yields at most second-order equations of motion~\cite{Horndeski}.
We shall therefore consider the effective limit of a fundamental theory which can be cast in the Horndeski action~\cite{kobayashi}
\begin{equation}
\label{eq:316}
    S = \integral \left[\frac{1}{2} \sum_{i=2}^5 \mathcal{L}_i(\g,\varphi)+\mathcal{L}_m(\g,\Psi_m)\right] \,,
\end{equation}
where the sum runs over the generalised Lagrangian densities
\begin{align}
    &\mathcal{L}_2 = G_2(\varphi,X) \,, \\
    &\mathcal{L}_3 = G_3(\varphi,X) \Box\phi \,, \\
    &\mathcal{L}_4 = G_4(\varphi,X) R + G_{4,X}(\varphi,X)\left[(\Box\varphi)^2+\varphi_{;\mu\nu} \varphi^{;\mu\nu}\right] \,, \\
    &\mathcal{L}_5 = G_5(\phi,X) G_{\mu\nu} \varphi^{;\mu\nu} - \frac{1}{6} G_{5,X}(\varphi,X)\left[(\Box\phi)^3 + 2\varphi_{;\mu}^\nu \varphi_{;\nu}^\alpha \varphi_{;\alpha}^\mu - 3\varphi_{;\mu\nu} \varphi^{;\mu\nu} \Box\varphi\right] \,.
\end{align}
The $G_i$'s are general functions of the field $\varphi$ and its kinetic term $X = -\,(1/2)\partial_{\mu}\varphi\, \partial^{\mu} \varphi$. Note that we recover the action~\eqref{eq:31} for the choices $G_2 = -V(\varphi)$, $G_4 = \varphi$ and $G_3=G_5=0$.

It is easy to see that the local self-tuning mechanism of Sec.~\ref{sec31} also operates in the Horndeski action.
The only thing needed is the additional coupling of $\varphi$ with the field strength of the three-form gauge field, introduced with the topological sector in Eq.~\eqref{eq:32}.
Since this term fixes the dynamics of $\varphi$ to take a constant value across the entire spacetime $\mathcal{M}$, all derivative terms in Eq.~\eqref{eq:316} vanish. The only remaining terms are $G_2(\varphi)$ and $G_4(\varphi)$.
With the freedom to redefine the scalar field as $\phi \equiv G_4$ and thus $G_2(\phi)=-V(\phi)$, one hence recovers the action~\eqref{eq:31}. Note that this can be generalised as well to Degenerate Higher-Order Derivative Scalar-Tensor (DHOST) \cite{langlois} theories beyond Horndeski gravity.

\subsection{Correspondence to global mechanism}\label{sec33}

We have found that the Einstein field equations~\eqref{eq:localeinstein} of the local self-tuning mechanism recover the field equations~\eqref{eq:27} of the global theory.
At the level of the action, we can also integrate out the topological sector of the local model, keeping in mind that it fixes the dynamics of the non-minimally coupled scalar field. This yields
\begin{equation}
\label{eq:313}
    S = \frac{\planck{2}}{2} \integral (R-2\Lambda) + \integral\mathcal{L}_m(\g,\Psi_m) + \sigma(\planck{2}) C \,,
\end{equation}
where we have set the constant $\varphi$ to $\planck{2}$ and $V=\planck{2}\Lambda$, and $C$ is the flux of the three-form gauge field that becomes a constant after integration.
Note that we do not have the last term in the global action~\eqref{eq:21}.
Let us therefore briefly explore its impact on the global self-tuning.

Variation of the action~\eqref{eq:313} with respect to $\planck{2}$ gives the constraint
\begin{equation}
    \frac{1}{2}\integral(R-2\Lambda) = -\sigma' C \,,
\end{equation}
where the prime denotes a derivative with respect to $\planck{2}$. Dividing both sides by the four-volume we have
\begin{equation}
    \frac{1}{2} \average{R} = \Lambda - \frac{\sigma' C}{\int d^4x \sqrt{-g}}
\end{equation}
and, hence, $\Lambda = \planck{-2}\langle T \rangle/2 - \tilde{C}$, where $\tilde{C} \equiv C \sigma'/\int d^4x \sqrt{-g}$.
Therefore, following the same computations as in Sec.~\ref{sec2}, one obtains in analogy to Eq.~\eqref{eq:27} the expression
\begin{equation}
    G_{\mu\nu}+\frac{1}{2-\alpha}\left[(1-\alpha)\Lambda+\frac{\planck{-2}}{2}\average{\tau} - \tilde{C} \right] \g = \planck{-2}\tau_{\mu\nu} \,,
\end{equation}
with the additional contribution $\tilde{C}$.
Now, $\tilde{C}$ can simply be absorbed into the free cosmological constant $\Lambda$.
Alternatively, one may consider the same arguments made in Sec.~\ref{sec2} for the vanishing of $\Delta \Lambda$, which with finite $C$ but infinite or large four-volume also motivate that $\tilde{C}$ should be vanishing.
Thus, from these considerations one can safely take the actions~\eqref{eq:21} and \eqref{eq:313} as describing the equivalent global self-tuning mechanism.

\section{Calculation of the residual $\Lambda$}\label{sec4}

We next inspect the space-time average $\average{\tau} = \integral \tau / \integral$,
where for simplicity we assume a matter-only universe, $\tau = \bar{\rho}_m$, with the total matter energy density $\bar{\rho}_m$ composed of baryonic and cold dark matter.
Note that we can safely neglect radiation components and the inflaton since the space-time integrals in $\average{\tau}$ are dominated by the late-time evolution~\cite{kalpa4}. Assuming a spatially perfectly homogeneous and isotropic background in $\Lambda$CDM for our cosmic manifold $\mathcal{M}$, it is easy to see that $\average{\tau}$ will vanish in a long-lived universe.
This is not a problem as for $\beta\neq\alpha$ we still have a free, radiatively stable, classical cosmological constant $\Lambda$ available in Eq.~\eqref{eq:localeinstein} (also see Eq.~\eqref{eq:27}).
In principle, $\Lambda$ could therefore simply be considered determined by measurement~\cite{kalpa1,lomb1}.
However, ideally we would also like to be able to understand the value of $\Lambda$ or at least understand why its fractional energy density $\Omega_{\Lambda}$ is comparable to the that of the total matter $\Omega_m$ today -- the \emph{Why Now?} problem.

Let us first consider the scenario $\alpha=\beta=1$~\cite{lomb1} such that the Einstein field equations are given by Eq.~\eqref{eq:26} and the residual cosmological constant is given by $\residual = \planck{-2}\average{\tau}/2$.
For Planck cosmological parameters~\cite{Aghanim:2018eyx}, this implies that the Universe should have undergone an immediate collapse at the scale factor $a = 0.926$, at an age of $0.88\,H_0^{-1}$, thus, about $1\, \mathrm{Gyr}$ in the past~\cite{lomb1}, and in contrast, an immediate collapse at the current epoch would account for 81\% of the observed value of the cosmological constant with a decreasing fraction for a longer future~\cite{lomb1}.
Similar values are also found for the global sequestering mechanism ($\alpha=\beta=0$)~\cite{lomb2}.
While the predicted value of the cosmological constant is interestingly close to measurement, it is not exact and moreover standard cosmology does not predict an imminent collapse of the Universe.

It was shown in Ref.~\cite{lomb2} that by an extension of the global sequestering mechanism the fact that the Cosmos is inhomogeneous on small scales can be used to bring the predicted value of $\residual$ into agreement with observations.
Thereby the cosmic matter content is split into isolated patches that ultimately form collapsed structures in finite time. This nonlinear evolution predicts $\Omega_{\Lambda}=0.697$ for the average $\residual$, which however fluctuates across the different patches.
In Ref.~\cite{lomb1} it was shown that in the global self-tuning mechanism with Eq.~\eqref{eq:26} ($\alpha=\beta=1$) the averaging over these maximally gravitationally bound structures leads to a prediction of $\Omega_{\Lambda}=0.704$ for all patches with their collapses occurring at some arbitrary time far into the future.
In order to realise the self-tuning of the residual cosmological constant to the observed value in both approaches the action must be extended with new sequestering terms or a nontrivial empty-space Lagrangian to prevent vacuum energy from gravitating or the residual from vanishing.

As we will show in the following, the approach conducted in Ref.~\cite{lomb1} can be significantly simplified and rendered very natural with no new terms required on top of Eqs.~\eqref{eq:31} and \eqref{eq:32} by simply adopting the observable Universe as the manifold $\mathcal{M}$ over which the integration in the action is performed.
The reason for this is that $\mathcal{M}$ itself will develop into a maximally gravitationally bound cell, where ultimately no observations can be made of any test objects residing outside of it.
For a long-lived Universe, the space-time integrals in $\average{\tau}$ are completely dominated by this future state of $\mathcal{M}$, which hence determines $\average{\tau}$.
More specifically, consider a spherical patch of physical size $R$.
In the Newtonian approximation its energy equation can be written as~\cite{Busha,dunner}
\begin{equation}
\label{eq:42}
    E = \frac{1}{2}\left(\frac{dR}{dt}\right)^2-\frac{G M}{R}-\frac{\Lambda_{\rm obs}}{6} R^2 \,,
\end{equation}
where $M$ is the total enclosed mass, $\Lambda_{\rm obs}$ is the observed cosmological constant driving the cosmic acceleration, and $E$ is the total energy per unit mass in the interior.
The critical shell of a patch, as the limit between expansion and collapse into the structure, is given for $dR/dt = 0$. It reaches a maximal value of
\begin{equation}
 \label{eq:43}
  R_{\mathrm{max}} \equiv
  \left(\frac{3GM}{\Lambda_{\rm obs}}\right)^{1/3}
\end{equation}
for gravitationally bound patches in the future of a $\Lambda$CDM universe. Assuming sphericity for simplicity, we now characterise the observable Universe $\mathcal{M}$ by its physical spatial radius $\xi(t)$, and the radius of the patch that will develop into the maximally gravitationally bound structure in the finite or infinite future of the Universe as $\zeta(t)$.
Hence, we have $\lim_{t \gg t_0}\zeta(t) = R_{\mathrm{max}}$ or $\lim_{t \rightarrow \infty}\zeta(t) = R_{\mathrm{max}}$, where $t_0$ denotes the current time.
Note that $\zeta(t)$ is not the same radius as $R(t)$.
We can now write
\begin{equation}
 \label{eq:avgtau1}
    \average{\tau} = \frac{\int_{\mathcal{M}}dV_4 \rho_m}{\int_{\mathcal{M}}dV_4} = \frac{\int dt \left[ \int_0^{\zeta(t)} dr \, r^2 \hat{\rho}_m + \int_{\zeta(t)}^{\xi(t)} dr \, r^2 \bar{\rho}_m \right] }{\int dt \left[ \int_0^{\zeta(t)} dr \, r^2 + \int_{\zeta(t)}^{\xi(t)} dr \, r^2 \right]} \,,
\end{equation}
where $\hat{\rho}_m$ and $\bar{\rho}_m$ denote the total matter density in the local matter patch and the cosmological background, respectively.
Note that we can compute $\hat{\rho}_m$ and $\zeta(t)$ using the spherical collapse model~\cite{lomb1,lomb2}.
Importantly, $\xi(t)\rightarrow\zeta(t)$ for $t\gg t_0$.
This is due to the accelerated background expansion, where in a finite time into the future everything outside of $\zeta(t)$ will be expelled out of the cosmic event horizon, and thus disappear from our detectors.
Moreover, any test object in the intermediate region between $R_{\rm max}$ and the event horizon will become unobservable as it will be exponentially redshifted away~\cite{krauss1,krauss2,adams}.
In the far future, the observable Universe around Earth will therefore reduce to the radius $\zeta(t)$.
Thus, the second integrals in the numerator and denominator of Eq.~\eqref{eq:avgtau1} are subdominant in a long-lived universe, for which we therefore find
\begin{equation}
    \average{\tau} \rightarrow \frac{\int dt \int_0^{\zeta(t)} dr  \, r^2 \hat{\rho}_m}{\int dt \int_0^{\zeta(t)} dr \, r^2}
    \rightarrow \frac{\int dt \int_0^{R_{\rm max}} dr  \, r^2 \hat{\rho}_m^{\rm max}}{\int dt \int_0^{R_{\rm max}} dr \, r^2} = \hat{\rho}_m^{\rm max} \,.
\end{equation}
Using Eq.~\eqref{eq:43}, we thus obtain
\begin{equation}
 \hat{\rho}_m^{\rm max} = \frac{3}{4\pi} \frac{M}{R_{\rm max}^3} = \frac{\Lambda_{\rm obs}}{4\pi G} \,,
\end{equation}
or in other terms,
\begin{equation}
 \label{eq:observedLambda}
 \planck{-2}\frac{\average{\tau}}{2} = \Lambda_{\rm obs} \,.
\end{equation}
We also confirm this solution numerically with the spherical collapse computations of Ref.~\cite{lomb1,lomb2}.
Hence, with Eq.~\eqref{eq:observedLambda} we find a self-consistent solution in Eq.~\eqref{eq:26}.
More generally, the classical cosmological constant in Eq.~\eqref{eq:localeinstein} becomes
\begin{equation}
 \planck{-2}\Delta V = \frac{(1-\alpha)\Lambda_{\rm obs} - \Delta\Lambda}{\beta-\alpha} \,.
\end{equation}
Note that Eq.~\eqref{eq:observedLambda} applies for $\average{\tau}$ independently of the self-tuning mechanism.
The same argument therefore also applies for the local sequestering mechanism, where Eq.~\eqref{eq:observedLambda} would determine $\Delta\Lambda$.
Interestingly, Eq.~\eqref{eq:observedLambda} is also found from considerations of the causal universe in Ref.~\cite{gaztanaga}.
It is worth emphasising as well that one also arrives at Eq.~\eqref{eq:observedLambda} considering a single test particle in empty space.
Hence, no additional terms in the actions \eqref{eq:31} and \eqref{eq:32} are required to cancel vacuum energy gravitation in empty space or to prevent $\Lambda_{\rm res}$ from vanishing (cf.~\cite{lomb1,lomb2}).

Finally, while we have found a self-consistent self-tuning mechanism that reproduces $\Lambda_{\rm obs}$ from the space-time average $\average{\tau}$, $\Lambda_{\rm obs}$ could still be of arbitrary value.
As was argued in Refs.~\cite{lomb1,lomb2}, however, the only relevant dynamical quantity in the determination of $\average{\tau}$ is the physical radius $\zeta(t)$ of the matter patch that evolves to become our local maximally gravitationally bound structure in the future of the Universe.
The \emph{Why Now?}~problem of the cosmological constant can therefore be phrased in terms of being located at a particular place in the evolution of $\zeta(t)$ such that
$\Omega_{\Lambda}(t_0) \sim \Omega_m(t_0)$ today, $t_0$, where $\Omega_{\Lambda}(t)\equiv\planck{-2}\bar{\rho}_m/(3H^2)$ and $\Omega_{\Lambda}(t)\equiv\Lambda_{\rm obs}/(3H^2)$ with $H$ denoting the Hubble function.
One can define the dimensionless physical top-hat radius $y(t)=\zeta(t)/a(t)/r_{\rm th}$, where $a(t)$ is the scale factor and $r_{\rm th}$ is the comoving radius of the top-hat overdensity that evolves into the maximally gravitationally bound structure, thus, $M=(4\pi/3)\bar{\rho}_{\rm m}(t_0) r_{\rm th}^3$ in Eq.~\eqref{eq:43}.
Adopting as the simplest ansatz a uniform prior on $y\in[0,1)$ to estimate our likely location in the evolution of $\zeta(t)$, we find the average expectation $y(t_0)=1/2$.
This expression can be solved for $t_0$ without assuming any values for the cosmological parameters~\cite{lomb1,lomb2}.
One then finds from this that $\Omega_{\Lambda}(t_0)=0.704$, in good agreement with observations~\cite{Aghanim:2018eyx}.
Instead of a flat prior on $y$, however, one may wish to construct a more physical prior, which likely involves the consideration of the evolution of stellar systems.
Star formation has peaked about 10 billion years in the past such that one may na\"ively expect a peak in the emergence of intelligent life about 5 billion years ago, assuming a similar biological evolution can be extrapolated from one sample.
Following the star formation history, the stellar formation has dropped by a factor of four by the time the Sun was formed, placing our existence at $t_0$ under these considerations not at the most likely location.
As was argued in Ref.~\cite{lomb2} considering instead a prior for stellar systems that contain heavier elements than iron, one may expect a shift of the peak of the relevant star formation history of about 5 billion years to later times to allow for the s-process to take place in typical stars, which would set our Sun close to the shifted peak position.
A cosmological peak for the emergence of intelligent life may then reasonably be expected close to $t_0$.
We leave a more detailed analysis of the likelihood of $\Omega_{\Lambda}(t_0)$ from such considerations to future work.

\section{Conclusions}\label{sec5}

Identifying the physical nature of the cosmological constant and the late-time accelerated expansion of our Universe is a prime endeavour to cosmology.
It is generally thought attributed to vacuum fluctuations.
However, quantum theoretical computations of this contribution to gravitational dynamics are off by several orders of magnitude.
Recently, a simple variation of the Planck mass in the Einstein-Hilbert action of GR in addition to the metric variation has been proposed as a remedy to this problem by introducing a self-tuning mechanism of the cosmological constant that prevents vacuum energy from fully gravitating.
Moreover, the evaluation of the resulting constraint equation under consideration of the evolution of the inhomogeneous cosmic matter distribution was shown to self-consistently reproduce the observed cosmological constant with an expected value for the current fractional energy density of $\Omega_{\Lambda}=0.704$, in good agreement with observations. Besides the non-gravitating vacuum energy, the global self-tuning mechanism therefore also explains the rise of the late-time accelerated cosmic expansion and the coincidence between the current energy densities of matter and the cosmological constant.

In this paper, we have developed a local theory from which the global self-tuning mechanism naturally emerges.
To achieve this, we have considered general scalar-tensor actions that can arise as the effective limit of a more fundamental theory with the additional presence of a topological sector in which the scalar field couples to the field strength of a three-form gauge field.
The resulting additional three-form field equations enforce constancy of the scalar field on the domain of the action, which reproduces the global self-tuning mechanism with the scalar field equation providing the constraint equation.
We then showed that the self-tuning mechanism provides a self-consistent framework that recovers the observed cosmological constant from the simple evaluation of the constraint equation over the observed Universe that in the future will reduce to the local maximally gravitationally bound structure.
This simplifies the previous picture where the constraint was evaluated on the inhomogeneous matter distribution with the employment of a nontrivial empty-space Lagrangian density to enable the self-tuning of a non-vanishing cosmological constant in empty space.
We discuss likelihood estimations for our location in the evolution of the local maximally gravitationally bound patch, finding that the observed value of the fractional energy density of the cosmological constant is in good agreement with expectations.
Finally, we lay out some ideas on how the local self-tuning mechanism can be used to absorb quantum gravity effects on the gravitational dynamics.
We leave a more detail analysis of that to future work.

\section*{Acknowledgments}

This work was conducted in the context of DSB's MSc thesis. LL acknowledges support by a Swiss National Science Foundation Professorship grant (No.~170547). Please contact the authors for access to research materials.

\appendix
\section{Generalised counter-term}\label{appendix1}

In Sec.~\ref{sec2}, we have for simplicity restricted our discussion of the global self-tuning mechanism to vacuum energy contributions that can be written as a power law of the Planck mass.
We have already provided a more general analysis for the local self-tuning formalism in Sec.~\ref{sec3}, allowing for an arbitrary dependence of $\Lambda_\vac$ on the Planck mass, where however for simplicity we have neglected the bare cosmological constant $\Lambda_B$.
We shall briefly discuss how an arbitrary dependence of $\Lambda_\vac$ on $\planck{2}$ is allowed in the global self-tuning mechanism and how an arbitrary $\Lambda_B$ is absorbed. We start from the action~\eqref{eq:21} and separate out the vacuum and bare components from the matter Lagrangian density, $\mathcal{L}(g^{\mu\nu},\Psi_m)=\Bar{\mathcal{L}}_m(g^{\mu\nu},\Psi_m)-(\Lambda_\vac+\Lambda_B)$, assuming $\Lambda_\vac$ and $\Lambda_B$ to be arbitrary functions of $\planck{2}$.
We now perform a similar separation for the classical cosmological constant, $\planck{2}\Lambda \rightarrow \planck{2}\Lambda+\Lambda_C+\Lambda_D$, where
$\Lambda$, $\Lambda_C$, and $\Lambda_D$ are arbitrary functions of $\planck{2}$. The action, hence, becomes
\begin{equation}
\label{eq:appendixa1}
    S = \integral\left[\frac{\planck{2}}{2}(R-2\Lambda) - (\Lambda_C+\Lambda_D+\Lambda_\vac+\Lambda_B) + \Bar{\mathcal{L}}_m(g^{\mu\nu},\Psi_m)\right] \,.
\end{equation}
Variation with respect to the metric gives the Einstein field equations
\begin{equation}
\label{eq:appendixa2}
    G_{\mu\nu} + \left[\Lambda + \planck{-2}(\Lambda_B+\Lambda_C+\Lambda_D+\Lambda_\vac)\right]\,g_{\mu\nu} = \planck{-2}\,\tau_{\mu\nu} \,.
\end{equation}
We shall parametrise the Planck mass dependence as
\begin{equation}
    \alpha_i \equiv \frac{\partial \ln{\Lambda_i}}{\partial \ln{\planck{2}}} \,,
\end{equation}
where the indices denote $i=\{B,C,D,\vac\}$ and we associate $\alpha$ to $\Lambda$ (or we could take $\beta=1+\alpha$ for $\planck{2}\Lambda$ in the notation of Sec.~\ref{sec3}).
Variation of the action~\eqref{eq:appendixa1} with respect to $\planck{2}$ yields the constraint 
\begin{equation}
\label{eq:appendixa5}
    (1-\alpha)\planck{2} \Lambda + (2-\alpha_\vac)\Lambda_\vac + (2-\alpha_B)\Lambda_B + (2-\alpha_C)\Lambda_C + (2-\alpha_D)\Lambda_D = \frac{\average{\tau}}{2} \,,
\end{equation}
where we have used the trace of Eq.~\eqref{eq:appendixa2}.
Finally, we solve this expression for the free function $\Lambda_C$ and introduce the result into the Einstein equations \eqref{eq:appendixa2} to get
\begin{eqnarray}
 \label{eq:einsteinlong}
    G_{\mu\nu}+\frac{1}{2-\alpha_C}\Bigg[(1 +\alpha - \alpha_C) \Lambda + \frac{\planck{-2}}{2} \average{\tau} + (\alpha_\vac-\alpha_C)\planck{-2} \Lambda_\vac & & \nonumber\\ 
    + (\alpha_B-\alpha_C)\planck{-2} \Lambda_B + (\alpha_D-\alpha_C)\planck{-2} \Lambda_D\Bigg] g_{\mu\nu} & = & \planck{-2} \tau_{\mu\nu} \,.
\end{eqnarray}
To cancel $\Lambda_\vac$ we therefore need $\alpha_C=\alpha_\vac$.
Note again that the constraint on $\Lambda_C$ only needs to apply for a given choice, or measurement, of the Planck mass, hence, $\Lambda_C$ does not change its explicit dependence on $\planck{2}$ (see Eq.~\eqref{constraint4}).
To prevent fine-tuning in $\alpha_C=\alpha_\vac$, we need $\Lambda_C\propto\Lambda_V$ with a proportionality factor that is independent of $\planck{2}$.
In particular, this recovers the power-law scenario in Sec.~\ref{sec2} with $\planck{2}\Lambda_\vac=\planck{2\alpha_{\vac}}\bar{\Lambda}_{\vac}$ and $\planck{2}\Lambda_C=\planck{2\alpha_{\vac}}\bar{\Lambda}_{\alpha}$.
The cancellation of the bare contribution then occurs straightforwardly if $\alpha_B=\alpha_\vac$.
Alternatively, the contribution cancels for
\begin{equation}
 \label{eq:LambdaDconstraint}
 \Lambda_D = \frac{\alpha_B - \alpha_\vac}{\alpha_\vac - \alpha_D}\Lambda_B \,.
\end{equation}
This corresponds to a fine-tuning of $\Lambda_D$, which is, however, not problematic since $\Lambda_B$ is not prone to radiative corrections.
Similarly, given the free classical cosmological constant $\Lambda$ in Eq.~\eqref{eq:einsteinlong}, $\Lambda_B$ can simply be absorbed into the choice of $\Lambda$.

\section{Graviton loops}
\label{appendix2}

It is well known that quantum corrections to gravity give contributions to the gravitational coupling and the cosmological constant.
In particular, the cosmological constant is modified by both 1PI matter and graviton loops.
The one-loop vacuum correction from the matter sector in curved space-time is given by~\cite{Martin:2012bt}
\begin{equation}
\label{eq:23}
    \Lambda_\vac = \planck{-2}\,\sum_i\,n_i\frac{m_i^4}{64\pi^2}\log{\left(\frac{m_i^2}{\mu_i^2}\right)} + \Lambda_\vac^{\mathrm{EW}} + \dots,
\end{equation}
where $i$ runs over the different particle species, $m_i$ denote their masses, $n_i$ represent their respective number of degrees of freedom with $+/-$ for bosons/fermions, and $\mu_i$ are unknown renormalization mass scales. The electro-weak phase transition contributes to $\Lambda_\vac$ as $\Lambda_\vac^{\mathrm{EW}} = -\planck{-2}\left(\sqrt{2}/16\right)\left(m^2_H/G_F\right)$, with Higgs boson mass $m_H$ and Fermi constant $G_F$. The ellipsis denotes further contributions, e.g., the QCD phase transition.
We have discussed how the one-loop correction and higher-order corrections are absorbed in the self-tuning mechanism in Secs.~\ref{sec2} and \ref{sec3}.
We shall now briefly discuss the graviton loops.

Generally, the vacuum and bare contributions to the cosmological constant arising from matter and graviton loops can be understood as some complicated function of the quadratic Planck mass.
We can perform the expansion
\begin{equation}
 \label{eq:vacuumcorrection}
    \Lambda_\vac(\planck{2}) = a_0 M^4 + a_1 \frac{M^6}{\planck{2}} + a_2\frac{M^8}{\planck{4}} + \dots = \sum^\infty_{n=0} a_n \frac{M^{4+2n}}{\left(\planck{2}\right)^n} \,,
\end{equation}
where $M$ is some renormalization mass scale.
Consider the expansion of a classical counter-term, $\Lambda_C = \sum_{m=-\infty}^\infty \bar{\Lambda}_m \left(\planck{2}\right)^m$.
For the two contributions to cancel we need $\partial\ln\Lambda_C/\partial\ln\planck{2} = \partial\ln\Lambda_\vac/\partial\ln\planck{2}$.
It is clear from this condition that we can only cancel off one arbitrary coefficient in Eq.~\eqref{eq:vacuumcorrection} as we run into a fine-tuning problem for the next coefficient (see Eq.~\eqref{eq:LambdaDconstraint} for an analogy).
But we can cancel off an overall scaling of each term, which is still an interesting property given that the expansion~\eqref{eq:vacuumcorrection} does not converge.
It should be furthermore emphasised, however, that graviton contributions have also been studied for the related sequestering mechanism in Ref.~\cite{kalpa5}.
A similar approach can be adopted for the cancellation of quantum gravity corrections in the self-tuning mechanism from the Planck mass variation presented here.
For instance, for quantum corrections with higher-derivative terms in Eq.~\eqref{eq:21} that are independent of Planck mass, there are no contributions to Eq.~\eqref{eq:25} or to the field equations, and if dependent on $\planck{2}$ by an overall power-law scaling of $\planck{2}$, they are cancelled by the same classical counter-term as in Sec.~\ref{sec2}.
It is also worth emphasising that in the scalar-tensor representation of Sec.~\ref{sec3}, a coupled Gauss-Bonnet invariant can be recast in Horndeski theory, and we have described how the self-tuning mechanism is operating for general Horndeski theories in Sec.~\ref{sec32}.
We leave a more detailed analysis of the effects of graviton loops on the self-tuning mechanism for future work.

\bibliographystyle{JHEP} 
\bibliography{refs} 

\end{document}